\documentclass[letterpaper]{article}%
\usepackage{amsmath}%
\usepackage{amsfonts}%
\usepackage{amssymb}%
\usepackage{graphicx}
\usepackage[draft]{hyperref}
\usepackage[margin=2.5cm]{geometry}
\usepackage[rgb]{xcolor}

\begin{document}

\title{\textbf{A photonic crystal cavity-optical fiber tip nanoparticle sensor for biomedical applications}}
\author{Gary Shambat,$^{1,}$\footnote{email: gshambat@stanford.edu}\hspace{1.5 mm} Sri Rajasekhar Kothapalli$^2$, Aman Khurana$^2$, J Provine$^1$,  
\\Tomas Sarmiento$^1$, Kai Cheng$^2$, Zhen Cheng$^2$, James Harris$^1$, 
\\Heike Daldrup-Link$^2$, Sanjiv Sam Gambhir$^{2,3}$, and Jelena Vu\v{c}kovi\'{c}$^1$
\\{\small $^1$Department of Electrical Engineering, Stanford University, Stanford, California 94305 USA}
\\{\small $^2$Molecular Imaging Program at Stanford, Department of Radiology and Bio-X Program,}
\\{\small Stanford University, Stanford, California 94305 USA}
\\{\small $^3$Department of Bioengineering, Department of Materials Science and Engineering,}
\\{\small Stanford University, Stanford, California 94305 USA} }

\date{}
\maketitle

\begin{abstract}
We present a sensor capable of detecting solution-based nanoparticles using an optical fiber tip functionalized with a photonic crystal cavity. When sensor tips are retracted from a nanoparticle solution after being submerged, we find that a combination of convective fluid forces and optically-induced trapping cause an aggregation of nanoparticles to form directly on cavity surfaces.  A simple readout of quantum dot photoluminescence coupled to the optical fiber shows that nanoparticle presence and concentration can be detected through modified cavity properties. Our sensor can detect both gold and iron oxide nanoparticles and can be utilized for molecular sensing applications in biomedicine. 
\end{abstract}

\section{Main text}

Nanoparticles (NPs) have recently been the subject of much attention for their uses in nanomedicine \cite{davis,peer}, molecular imaging \cite{ferrari, cao}, and phototherapy \cite{hirsch, huang}. Due to their small sizes (typically less than 100 nm), nanoparticles can infiltrate cancerous tissue through the enhanced permeability of the vasculature and can act as markers for imaging tumors \cite{qian}. Active targeting of nanoparticles to specific tissue or cell surface proteins can be accomplished through proper ligand chemistry. Gold NPs, for example, can be detected through wide-field imaging of covalently linked Raman reporter molecules and have been used to image small tumors in mice models \cite{qian, mohs, keren}. Plasmonic gold nanoshells on the other hand have been used for photothermal ablation of tumors by locally heating tissue with a laser pump \cite{hirsch}.   

Recently we developed a method to functionalize optical fiber tips with semiconductor photonic crystal (PC) cavities by using a simple epoxy transfer process \cite{shambatfiberPC}. With this design, light can be coupled back and forth between the cavities and the optical fibers for efficient optical readout that avoids a bulky free-space setup. Furthermore, the optical fiber tip is well suited for remote sensing measurements in tough environments such as the body due to its compact form factor. In this letter, we show how our fiber photonic crystal, or fiberPC, can operate as a nanoparticle sensor for both gold and iron oxide and can even quantitatively determine NP concentration. In contrast to previous wide-field imaging techniques requiring high (20-200mW) pump powers and bulky optics \cite{qian, mohs, keren}, our method is fully embedded with the fiber which can be envisioned as an endoscopic tool that requires less than 1 mW. With this distinct modality, nanoparticles can be detected simply by their proximity to the cavity, allowing for a unique avenue of NP sensing (e.g., intraoperatively) in future biomedical studies.    

Devices were fabricated following our previous epoxy transfer methodology \cite{shambatfiberPC}. Photonic crystal cavities were made out of a 220 nm thick membrane of GaAs with embedded high density InAs quantum dots (QDs) as internal light sources. Modified L3 defect cavities were patterned with a lattice constant a = 330 nm and hole radius r = 0.22a \cite{akahane}. In contrast to our former work, we generate just a single cavity at the center of our circular template rather than an array of many cavities. This is because we have improved our assembly process to have minimal alignment error (0-3 $\mu$m typical lateral offset) and because monitoring just a single cavity is preferable for sensing. Fig. 1(a) shows a scanning electron microscope (SEM) of a completed fiberPC device. As in our previous work, the membrane matches and smoothly covers the optical fiber tip with epoxy regions located away from the PC cavity center.       
      
A schematic of our test setup is shown in Fig. 1(b). The fiberPC pigtail is spliced to a patch cable and connected to a custom made wavelength division multiplexor (WDM) that is built to combine 1300 nm and 830 nm signals (Micro-Optics Inc.). We pump our devices with an 830 nm laser diode (LD) and collect the return photoluminescence (PL) with our spectrometer as we insert our sensors into various solutions. Laser pump powers ranged from 10 $\mu$W to 2.5 mW (measured prior to the fiberPC membrane) and integration times were typically several hundred ms. The brightness of our signals confirms the good quality of our device design as well as the high efficiency of collection. For our first experiments we use solutions of 15 nm gold nanoparticles that are colloid stabilized with carboxylic acid terminated polyethylene glycol. 

Fig. 2(a) shows a PL spectrum of a fiberPC device in air before any testing when pumped with 250 $\mu$W of laser power. A fundamental cavity mode appears at 1278 nm along with several other peaks at longer wavelength which are likely PC band edge modes. The Q-factor of the fundamental mode is 800 prior to solution testing. We next insert our device into a 12.5 nM solution of gold nanoparticles as seen in Fig. 2(b). Inside this solution the cavity mode redshifts by 12 nm, increases in quality factor to 1330, and decreases in emission intensity, all a result of the higher refractive index cladding provided by the water (versus the original air). Incidentally, the Q-factor goes up in this case due to the improved symmetry of the oxide and water claddings which reduces lossy TE-TM modal conversion \cite{shambaterb}. At this stage, metal nanoparticles are not detected since a control solution of water only has the exact same effect on our device. When we retract our device from the solution with the pump laser turned off during the retraction we find that the PL spectrum (observed 10 seconds after retraction with the pump then turned on) replicates the original air spectrum prior to solution testing (Fig. 2(c)). An optical microscope image of the device shows that the cavity region is not modified and there is a circular coffee-ring like annulus where a small amount of NPs deposited as a droplet evaporated from the fiber tip \cite{bhardwaj} (see Supplementary material for a real-time video of a droplet evaporating on a fiber tip). 

A very interesting and different scenario results when we keep our laser pump on during the fiberPC sensor withdrawl from the solution (see Supplementary material for a real-time video of a droplet evaporating with the laser turned on). As shown in the PL spectrum from Fig. 2(d), the cavity modes are now almost completely eliminated. The original fundamental mode is only barely visible now at 1298 nm, or 20 nm redshifted from air, and the Q-factor has dropped to 650. Examining the optical microscope picture it is clear that a large aggregation of metal NPs has formed directly over the cavity, suggesting that these nanoparticles are responsible for the change in cavity parameters. Likely, NPs cause some absorption of the optical field as well as perturb the uniform refractive index of the cavity cladding resulting in excess scattering \cite{kubo}. The higher refractive index of the particles result in a large cavity redshift as well. SEM close-up images of the metal aggregate are shown in Fig. 3(a).  

An explanation of these results requires an accurate understanding of the numerous physical mechanisms at play here. Recent studies of nanoparticle aggregation in evaporating droplet systems both with and without optical illumination have shown that convection, hydrothermal effects, surface forces, and chemical interactions all take place \cite{bahns, bhardwaj, kang}. Our proposed model is visualized in Fig. 3(b). When a fiberPC sensor is retracted from a NP solution, a nanoliter sized droplet is formed on the fiber tip. This droplet begins to quickly evaporate in air, which causes vigorous convection currents within the droplet \cite{bahns, bhardwaj}. Assisting in these evaporative convection currents is a hydrothermal contribution from the pump laser \cite{bahns}. Although the pump laser is meant to provide excitation for just the semiconductor cavity, we find that 55\% of the light is transmitted beyond the membrane. This weakly focused light will be absorbed by both the water and nanoparticles resulting in a temperature rise in the center portion of the droplet of at most a few 10s of K (for our laser intensities) \cite{bahns}. Such a small temperature rise can still have a large impact on enhancing the Marangoni convective currents that circulate in a toroidal pattern, propelling NPs into the center of the droplet \cite{bahns, bhardwaj}. As for the QD photoluminescence, we don't believe it plays a significant role in NP aggregation since it is orders of magnitude weaker than our pump laser.

The precise manner of NP aggregation is not quite as well understood in the literature, but several mechanisms are possible. Bahns et al. concluded that high temperature rises in excess of 100K were responsible for carbon to metal NP wetting interactions \cite{bahns}. However, our temperature rises are far too low to allow for gold-gold interactions. Pure optical trapping via the gradient optical force as observed by Yoshikawa et al. is ruled out here since our optical field is orders of magnitude too low for this \cite{yoshikawa}. Instead we believe that light driven photochemical interactions are likely responsible for NP aggregation. Laser light could potentially remove repulsive capping ions from NP surfaces or could even cause ligands to dissociate from the NPs, allowing nanoparticle-nanoparticle Van der Waals attractive interactions \cite{bahns, zhang}. More detailed studies need to be done to elucidate the precise attractive mechanisms. We also find that the aggregation process is reversible, and that a subsequent dip of the fiberPC tip in water or in the original NP solution with the pump laser turned off results in a clean washing of the NPs. Therefore the sensors are not limited to a one-time use. 

As further evidence that droplet evaporation and convection effects play an important role, we repeat the experiment with a bare fiber having no semiconductor membrane (see Supplementary material for a video of a droplet evaporating on a bare fiber tip). We observe a drastically weaker NP aggregation effect on these bare fiber tips. The contact angles measured on silica and GaAs surfaces were 10 and 50 degrees, respectively, suggesting that the much smaller droplet on silica surfaces like the bare fiber tip prevent effective convection of NPs into a central aggregate \cite{uno}. Therefore not only does a PC cavity provide optical feedback for a sensing measurement, but the semiconductor surface itself is a necessary component for proper NP concentration.  

We next look at the power and concentration dependencies of gold NP aggregation on fiberPC sensors. Fig. 4(a) shows a curve of the wavelength redshift for the same device as in Fig. 2 but now in a very low concentration solution of 0.8 nM gold NPs. Unsurprisingly the curve is monotonically increasing in wavelength shift with pump power, suggesting that laser driven convection and aggregation are enhanced with higher power. Meanwhile, Fig. 4(b) shows the wavelength shift of another fiberPC device when the pump power is held constant at our laser diode's maximum of 2.45 mW as the NP concentration is varied. We see a nearly linear dependence of wavelength shift with NP concentration, indicating that our sensor can be used to quantitatively measure NP concentration. Our detection limit for this device is at 100 pM, which corresponds to a 0.7 nm wavelength shift (equal to the cavity half width); however, we believe the detection limit could easily be improved by using higher pump powers or cavities with higher Q-factors. In Fig. 4(c) we examine the other limit of a highly concentrated NP solution (25 nM) pumped at an extremely low power of only 12 $\mu$W. Even for this very small pump power, we observe a significant redshift of 15.2 nm and a Q-factor degradation from 1050 to 650. Judging from the optical microscope image in Fig. 4(d) alone, one would conclude that no NPs had been detected; however, the fiberPC is sensitive to even miniscule aggregations of NPs that are not patently visible with the eye.

Finally, we demonstrate the versatility of our sensor by switching detection to iron oxide NPs. We investigated the ultrasmall superparamagnetic iron oxide compound ferumoxytol (Feraheme, AMAG Pharmaceuticals Inc.). Ferumoxytol is an FDA-approved iron supplement that has been used in patients for intravenous treatment of iron deficiency anemia \cite{schwenk}. Due to its superparamagnetic properties, ferumoxytol has also been used as a magnetic resonance (MR) contrast agent \cite{castaneda}. Ferumoxytol NPs have a core diameter of 7 nm and are coated with carboxymethyl dextran for colloid stabilization. Fig. 5(a) and (b) show PL spectra of a fiberPC sensor before and after submersion into a 400 $\mu$g/mL (or 533 nM) concentration solution of ferumoxytol when pumped at 1.75 mW. We chose this concentration because it is the value used when labeling cells for MR experiments \cite{castaneda}. As seen in Fig. 5, the cavity peaks once again redshift, this time by 21.8 nm and the Q-factor of the right-most mode decreases from 1000 to 770. The optical microscope picture in Fig. 5(b) shows a very similar aggregation at the center, this time from ferumoxytol NPs rather than gold NPs. A close-up SEM in Fig. 5(c) clearly shows the NP aggregation smothering the cavity. Contours of the aggregation highlight the fluidic deposition of the NPs much like wet sand dropped on a surface. We speculate that similar physical processes of convective concentration and photochemical binding are responsible for these NP aggregations.

In summary, we have demonstrated a nanoparticle sensor using a semiconductor photonic crystal cavity-optical fiber tip device. The cavity-on-a-fiber platform provides robust optical feedback which can be used to sense changes in its external environment through wavelength, Q-factor, and intensity information. Nanoparticle concentrations can be quantitatively determined based on the cavity wavelength shift as well. In contrast to a sensor built on a large chip substrate, integration of a sensor on a fiber tip with fascile measurement optics could allow for remote testing in difficult environments such as in the human body. This modality of sensing could be applied to various areas in biomedicine and nanoscience and is likely to work for numerous other nanoparticles commonly found in research. 

\section{Acknowledgements}
Gary Shambat is supported by the Stanford Graduate Fellowship. Gary Shambat also acknowledges the NSF GRFP for support. The authors acknowledge the financial support from NCI ICMIC P50CA114747 (SSG),  NCI CCNE-TR U54 CA119367 (SSG), and CCNE-T U54 U54CA151459 (SSG). Work was performed in part at the Stanford Nanofabrication Facility of NNIN supported by the National Science Foundation. We also thank Kelley Rivoire for helpful discussions.

\section{List of captions}

FIG 1. \textbf{(a)} Tilted SEM images of a completed GaAs fiberPC device. The two small blurry circles are where epoxy has been applied. The circular semiconductor template consists of an outer release region surrounding a central photonic crystal of size 20 x 25 $\mu$m. At the center of the PC is an L3 defect cavity as seen in the second inset. \textbf{(b)}  Schematic of the optical setup for measuring nanoparticles. Pump light from a laser diode is sent to the fiberPC tip through a custom 830/1300 nm WDM where it is absorbed above band by the GaAs semiconductor. Internal QDs embedded in the GaAs membrane emit PL both outward from the device and back into the fiberPC core where it can be subsequently detected by a spectrometer. 
\\
\\
FIG. 2. \textbf{(a)} PL spectrum of a fiberPC device in air and prior to testing when pumped at 250 $\mu$W. Optical image on the right shows the fiber tip face before testing. Arrows indicate epoxy droplets. \textbf{(b)} PL spectrum of the same sensor now in a 12.5 nM gold NP solution. The cavity fundamental mode has redshifted, increased in quality factor, and dropped in intensity. \textbf{(c)} PL spectrum in air again after the device has been retracted with the pump laser turned off during retraction. The spectrum is almost identical to that in \textbf{(a)} and the fiber tip image shows only a slight circular deposition of NPs on the outer rim. \textbf{(d)} PL spectrum in air of the same device but after having retracted the fiber tip with the laser pump turned on. The cavity modes are almost completely removed from the spectrum and a large circular aggregation of nanoparticles (indicated by the arrow) is seen in the microscope image.
\\
\\
FIG. 3. \textbf{(a)} SEM images of a fiberPC sensor with a metal NP aggregation on the cavity. Inset shows a close-up image. \textbf{(b)} Schematic model of the nanoparticle aggregation effect. A solution droplet on the fiber tip begins to evaporate as indicated by the outward flowing arrows (the contact angle of the droplet here is exaggerated for clarity). Meanwhile, part of the optical pump transmits through the thin photonic crystal membrane. Water and nanoparticles in the weak focus absorb the pump laser light, raising the local temperature of the water and setting up hydrothermal gradients. Combined with evaporation, the hydrothermal gradients create Marangoni convective flow which circulates fluid in a toroidal pattern (shown by the circles). This circulation propels NPs into the center of the droplet where they begin to aggregate most likely due to photochemical processes. Eventually, all the water in solution evaporates and only a deposition of NPs on the cavity remains. During this whole process, the quantum dot PL back-coupled into the fiber is observed with a spectrometer. 
\\
\\
FIG. 4. \textbf{(a)} Pump-power dependent wavelength shift of the fiberPC sensor from Fig. 2 now in a 0.8 nM solution of metal NPs. \textbf{(b)} Concentration dependent wavelength shift of a different sensor when the pump power is held at 2.45 mW. \textbf{(c)} PL spectrum of a different device before testing and associated optical microscope image. \textbf{(d)} PL spectrum of the same device as in \textbf{(c)} after retraction with a 12 $\mu$W pump laser turned on in a 25 nM metal NP solution and associated image. 
\\
\\
FIG. 5. \textbf{(a)} PL spectrum of a fiberPC device and the associated optical image of the fiber face prior to testing with ferumoxytol.  \textbf{(b)} PL spectrum of the sensor after retraction from a 400 $\mu$g/mL solution of ferumoxytol with a 1.75 mW pump. The spectrum changes are similar to those caused by metal NPs with a large redshift, reduction in Q-factor, and reduction in peak intensity. The optical image shows a circular aggregate at the fiber center as seen previously. \textbf{(c)} Close-up SEM image of a ferumoxytol aggregate. The scale bar is 2 $\mu$m. 

\newpage

\begin{figure}[htp]
\centering
\includegraphics{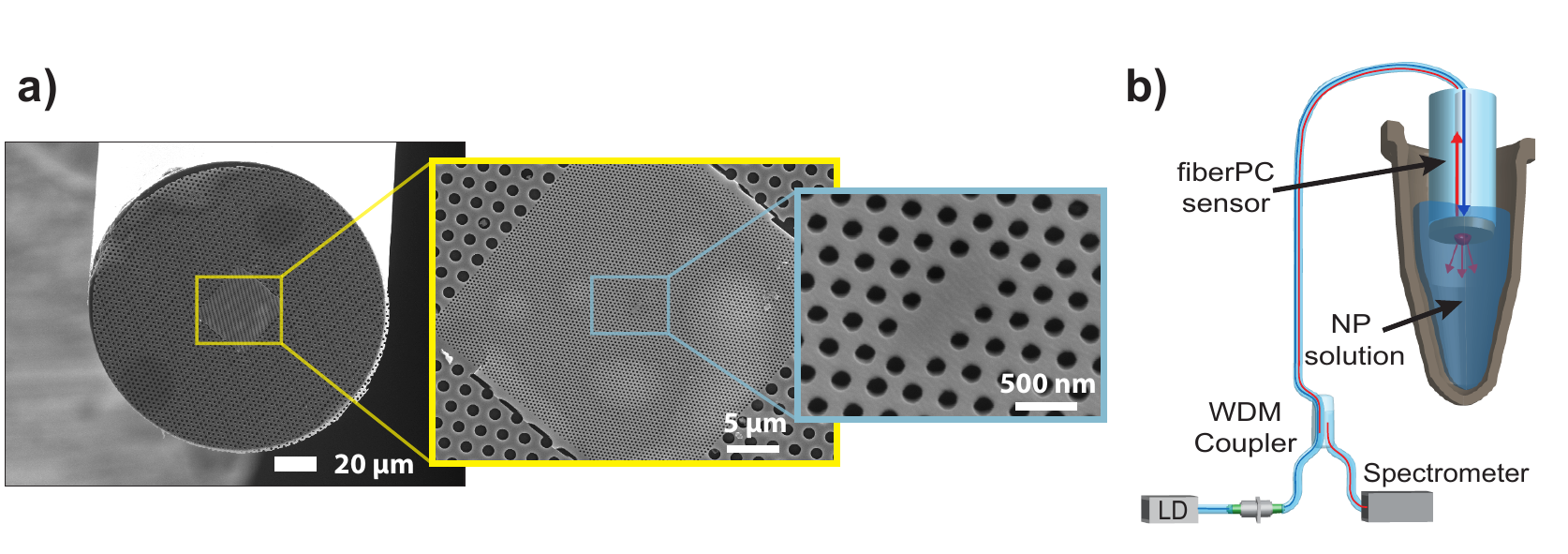}
\end{figure}

\newpage

\begin{figure}[htp]
\centering
\includegraphics{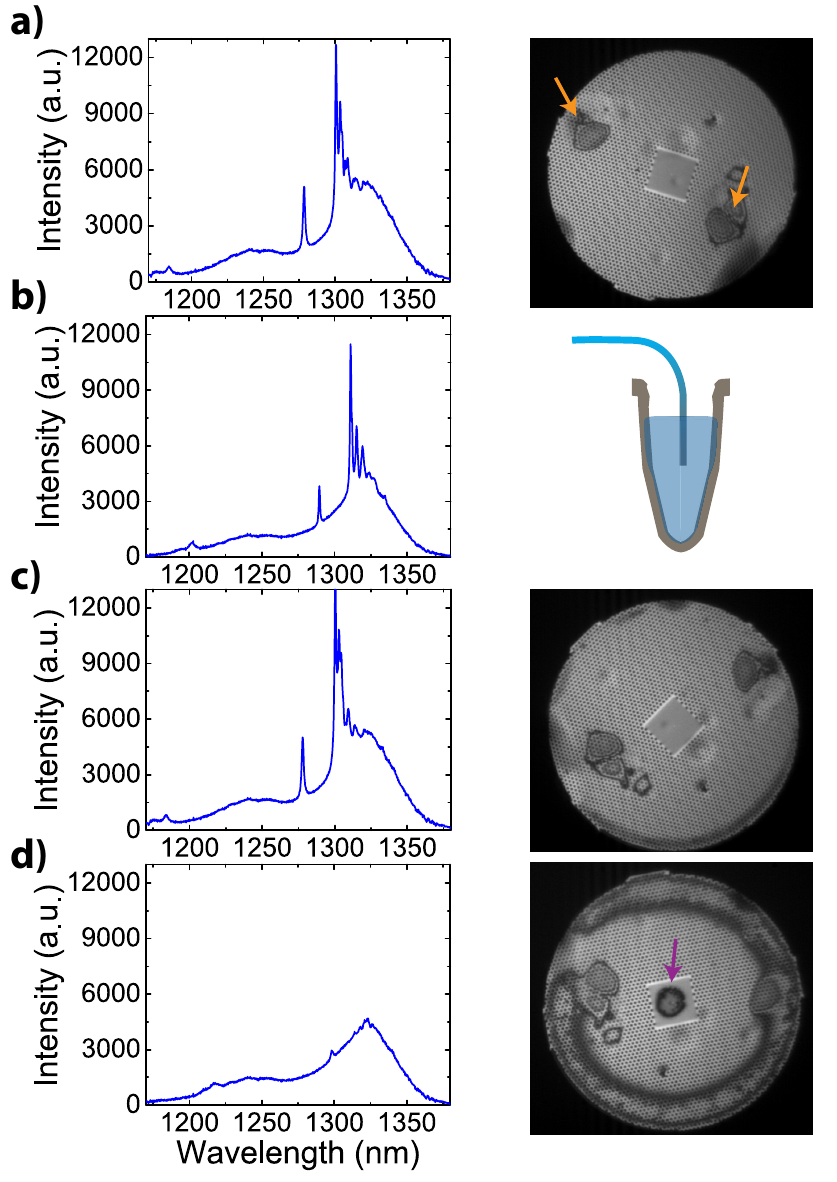}
\end{figure}

\newpage

\begin{figure}[htp]
\centering
\includegraphics{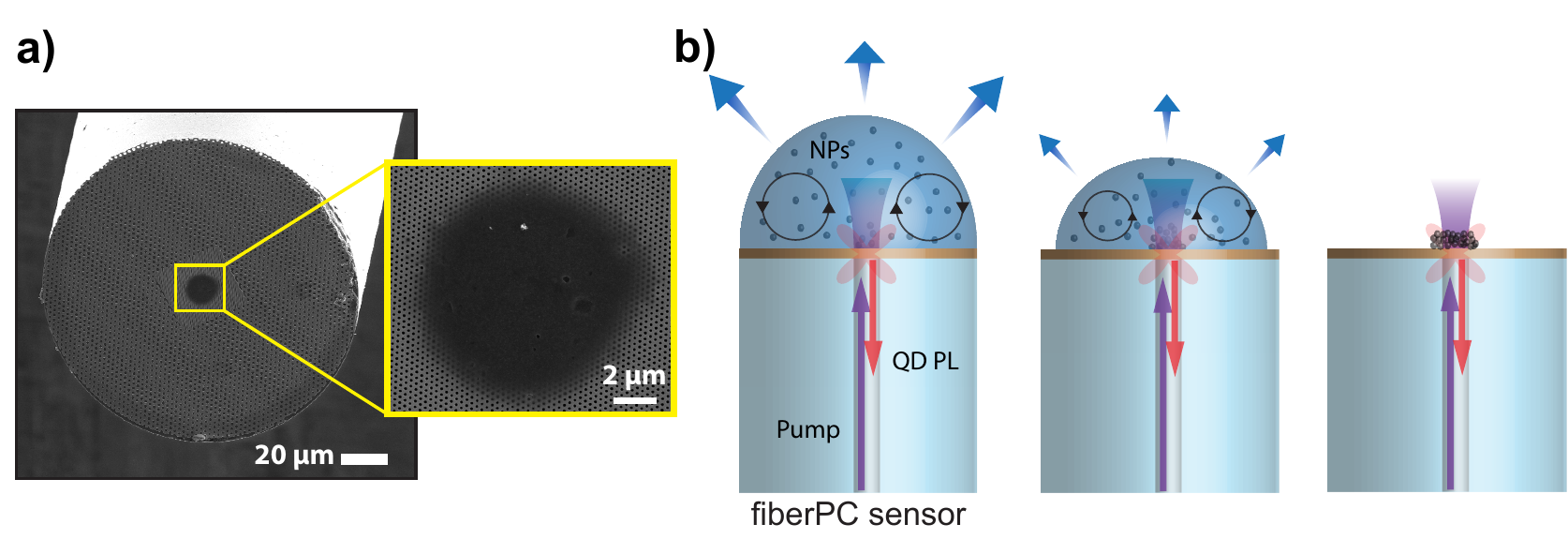}
\end{figure}

\newpage

\begin{figure}[htp]
\centering
\includegraphics{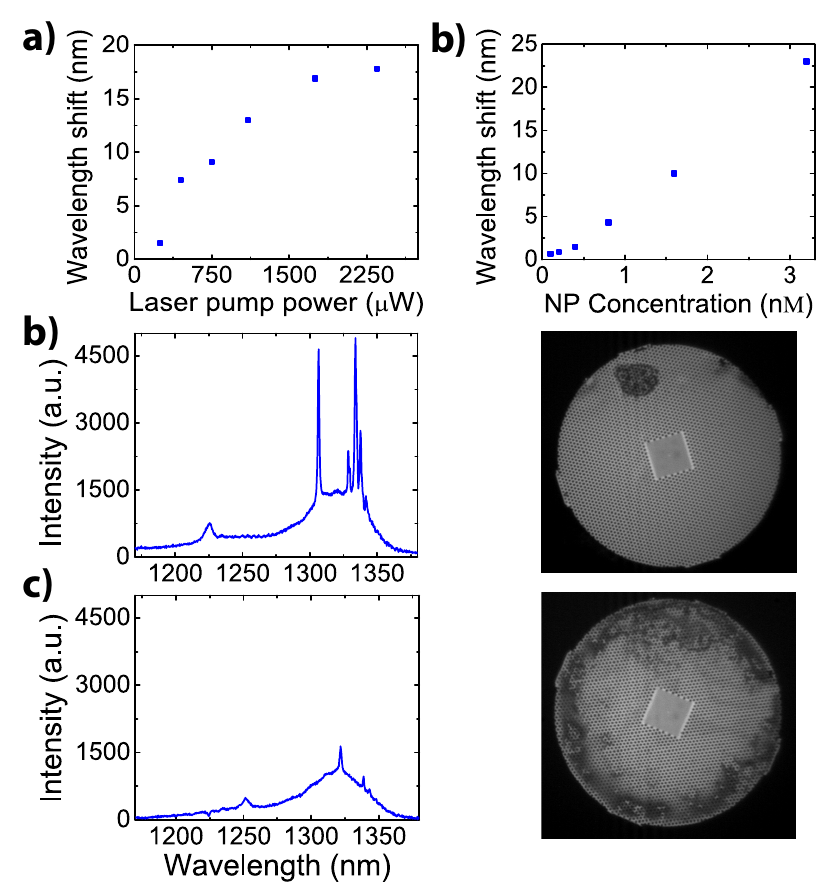}
\end{figure}

\newpage

\begin{figure}[htp]
\centering
\includegraphics{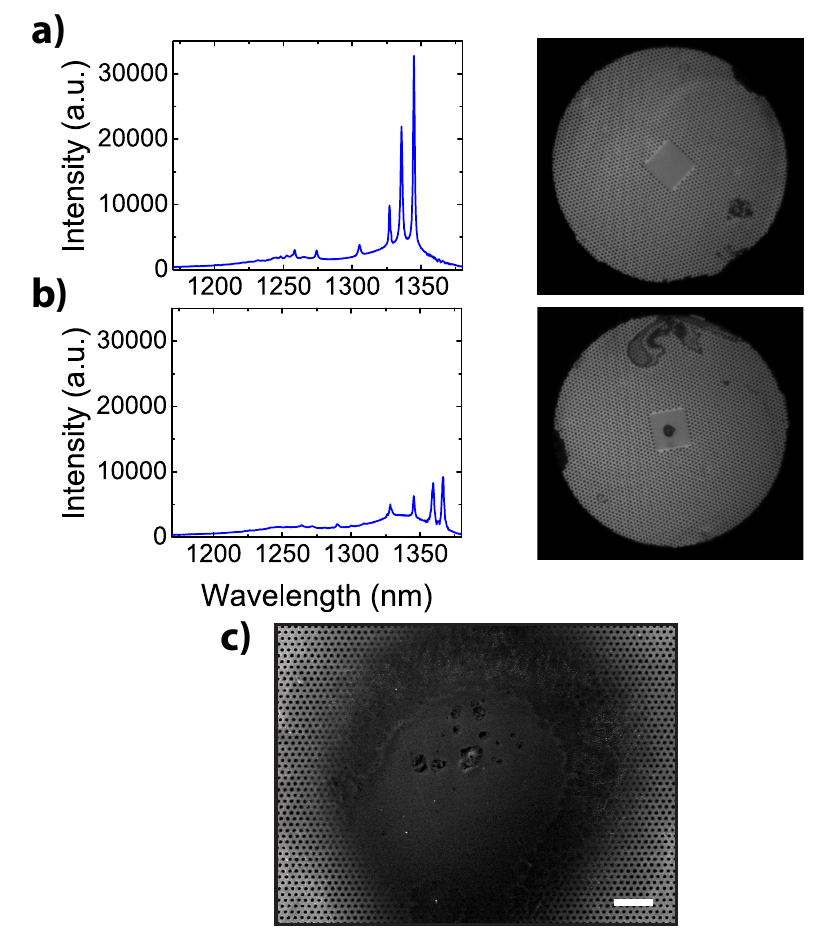}
\end{figure}

\end{document}